\documentclass{aa}  
\usepackage{txfonts}
\usepackage{newtxtext,newtxmath}
\usepackage[T1]{fontenc}
\usepackage{bm}
\usepackage{booktabs}
\usepackage{array}
\usepackage{enumerate}
\usepackage{ulem}
\usepackage{threeparttable}
\usepackage{multirow}
\usepackage{soul}
\setstcolor{red}
\usepackage{color}
\usepackage{graphicx}	
\usepackage{amsmath}	
\usepackage[colorlinks=true,linkcolor=blue,citecolor=blue,urlcolor=blue]{hyperref} 
\usepackage{extarrows}
\usepackage{amssymb}
\usepackage{hyperref} 
\usepackage{orcidlink}
\usepackage[switch]{lineno}

\DeclareRobustCommand{\VAN}[3]{#2}
\let\VANthebibliography\thebibliography
\def\thebibliography{\DeclareRobustCommand{\VAN}[3]{##3}\VANthebibliography}

\begin{document} 

\title{Shared Properties of Merger-driven Long-duration Gamma-Ray Bursts}


\author{Yacheng Kang\inst{1, 2, 3,\thanks{email: yckang@stu.pku.edu.cn}}\orcidlink{0000-0001-7402-4927}
		\and
		Jin-Ping Zhu\inst{4, 5}\orcidlink{0000-0002-9195-4904}
		\and
		Yu-Han Yang\inst{3}\orcidlink{0000-0003-0691-6688}
		\and
		Ziming Wang\inst{1, 2}\orcidlink{0000-0002-8742-8397}
		\and
		Eleonora Troja\inst{3}\orcidlink{0000-0002-1869-7817}
		\and\\
		Bing Zhang\inst{6, 7}\orcidlink{0000-0002-9725-2524}
		\and
		Lĳing Shao\inst{2, 8,\thanks{email: lshao@pku.edu.cn}}\orcidlink{0000-0002-1334-8853}
		\and
		Zhuo Li\inst{1, 2}\orcidlink{0009-0003-0571-0669}}
		

\institute{Department of Astronomy, School of Physics, 
		   Peking University, Beĳing 100871, China
           \and
           Kavli Institute for Astronomy and Astrophysics, 
           Peking University, Beĳing 100871, China
           \and
           Department of Physics, University of Rome ``Tor Vergata'', 
           via della Ricerca Scientifica 1, 00100, Rome, Italy
           \and
           School of Physics and Astronomy, Monash University, Clayton,
           VIC 3800, Australia
           \and
           OzGrav: The ARC Centre of Excellence for Gravitational Wave
           Discovery, Australia
           \and
           Nevada Center for Astrophysics, University of Nevada, Las
           Vegas, NV 89154, USA
           \and
           Department of Physics and Astronomy, 
           University of Nevada, Las Vegas, NV 89154, USA
           \and
           National Astronomical Observatories, 
           Chinese Academy of Sciences, Beĳing 100012, China}


\date{Received June 1, 2024 / Accepted June 1, 2024}


\abstract
{The recent detections of bright optical/infrared kilonova signals following two long-duration gamma-ray bursts (LGRBs), GRB\,211211A and GRB\,230307A, have significantly challenged the traditional classification of GRBs. These merger-driven LGRBs may represent a distinct GRB population, sparking interests in their progenitors and central engines.}
{Since traditional GRB classification methods often struggle to distinguish merger-driven LGRBs from traditional merger-driven short-duration GRBs resulting from compact object mergers and collapse-driven LGRBs produced by massive stars, this work aims to explore the shared properties in terms of hardness, energy, and duration among currently observed merger-driven LGRB events, thereby identifying their observed differences from the traditional GRB population.}
{We collect a sample of merger-driven LGRBs with known redshifts, including observed information on their main emission (ME) and whole emission (WE) phases. Treating ME and WE properties as two independent sets of information, we apply several GRB classification methodologies to explore their potential shared properties.}
{Using the phenomenologically defined Energy-Hardness ($EH$) parameter, characterized by the intrinsic hardness and energy of GRBs, and the duration of GRBs, we identify a probable universal linear correlation across merger-driven LGRBs, regardless of whether their ME or WE phases are considered.}
{We propose that such shared properties of merger-driven LGRBs are unlikely to arise from the low-redshift selection effect and they become particularly intriguing when compared with the relatively weak correlations or lack of correlation observed in traditional merger-driven short-duration GRBs (with or without extended emissions) and collapse-driven LGRBs. Our newly proposed correlation highlights the necessity for further investigation into the observations of merger-driven LGRBs and the physical mechanisms underlying the empirical correlation.}


\maketitle


\section{Introduction}
\label{ sec:intro }

Gamma-ray bursts (GRBs) are the most luminous explosions in the Universe, characterized by the emission of one or multiple $\gamma$-rays flashes. Statistical analysis of their durations reveals a bimodal distribution with a separation at $\sim2\,\mathrm{s}$, indicating the existence of two distinct progenitor classes \citep{Kouveliotou:1993yx}.\footnote{Specifically, their bimodal lognormal distributions have mean values of $\simeq 0.5\,\mathrm{s}$ for short-duration GRBs and $\simeq 30\,\mathrm{s}$ for long-duration ones \citep{Kouveliotou:1993yx, 1994MNRAS.271..662M}. The overlap between the two distributions challenges a clear-cut distinction.} Long-duration GRBs (LGRBs), with durations $\gtrsim2\,\mathrm{s}$, are generally attributed to the core collapse of massive stars \citep{Woosley:1993wj, Paczynski:1997yg, Woosley:1998hk, MacFadyen:1998vz}, supported by their associations with broad-lined Type Ic supernovae \citep[SNe;][]{Galama:1998ea, Woosley:2006fn}. Short-duration GRBs (SGRBs), lasting $\lesssim2\,\mathrm{s}$, are widely believed to originate from mergers of compact binary systems, such as neutron star--neutron star (NS-NS) and black hole--neutron star (BH-NS) systems \citep{Paczynski:1986px, Eichler:1989ve, 1991AcA....41..257P, Narayan:1992iy}. The landmark gravitational-wave event GW170817 \citep{LIGOScientific:2017vwq}, accompanied by electromagnetic counterparts, GRB\,170817A \citep{LIGOScientific:2017zic, Goldstein:2017mmi, Savchenko:2017ffs, 2018NatCo...9..447Z} and AT\,2017gfo \citep{Coulter:2017wya, Evans:2017mmy, Pian:2017gtc, Kilpatrick:2017mhz, Troja:2017nqp}, provided smoking-gun evidence for SGRBs and kilonovae (KNe) originating from NS-NS mergers \citep{Li:1998bw, Metzger:2010sy}. This milestone co-detection marked a breakthrough in multimessenger astrophysics and is widely regarded as a watershed moment in both astronomy and physics. The connection between gravitational-wave BH-NS mergers and SGRBs remains to be confirmed, since the gravitational-wave BH-NS events observed so far are possibly plunging events \citep[e.g.,][]{Abbott2021:bhns,Zhu2021,Zhu2024,Abac2024:bhns}.

However, using duration alone for GRB classification can sometimes introduce ambiguity. For example, a notable subset of GRBs exhibits a short-hard main emission (ME) phase, similar to normal merger-origin SGRBs, but it is followed by an extended emission (EE) component that lasts for tens of seconds and features a softer energy spectrum, which could be wrongly identified as collapse-driven LGRBs based on their overall properties \citep{Lazzati:2001gp, Connaughton:2001iz, Montanari:2005nb, Norris:2006rw, Norris:2009rg, Gao:2016uwi}. Additionally, the detection of optical/infrared KN signals following certain LGRBs with ME durations longer than $2\,\mathrm{s}$ challenges traditional GRB classifications, notably in GRB\,060614 \citep{Gehrels:2006tk, DellaValle:2006nf, Gal-Yam:2006qle, Zhang:2006mb, Yang:2015pha, Jin:2015txa}, GRB\,211211A \citep{Rastinejad:2022zbg, Yang:2022qmy, Troja:2022yya, 2022ApJ...933L..22Z}, and GRB\,230307A \citep{JWST:2023jqa, Sun:2023rbr, 2024Natur.626..742Y}. These observations suggest that a group of LGRBs should have merger origins. Intriguingly, such long durations are unexpected for merger events, as the accretion timescales required for jet launching are generally thought to be relatively short \citep{Narayan:1992iy}. Consequently, merger-driven LGRBs may represent a distinct GRB population, sparking interest in models where the progenitors and central engines of such events are unassociated with the core-collapse of massive stars \citep{Gao2022,Zhu:2022kbt, Gompertz:2022jsg, Zhong:2023zwh, Dichiara:2023goh, Gottlieb:2023sja, Gottlieb:2024mwu, Wang:2024ijo, Du:2024kst, Zhong:2024alw, Peng:2021knv, Chen:2024arq, Meng:2023wpf, Zhang:2024syc}.

Since commonly used GRB classification methods often struggle to distinguish merger-driven LGRBs from traditional merger-driven SGRBs and collapse-driven LGRBs, this work aims to investigate the shared properties in terms of hardness, energy, and duration among currently observed merger-driven LGRBs, as well as to identify how they differ from these traditional GRBs. The organization of this paper is as follows. We first present the construction of GRB samples in Section~\ref{ sec:sample }. In Section~\ref{ sec:corre }, we report our results and detailed analyses on different classification methods for merger-driven LGRBs, followed by a detailed discussion of our newly proposed relations in Section~\ref{ sec:discu }. Finally, Section~\ref{ sec:conclu } concludes the paper. We adopt a standard ${\rm \Lambda CDM}$ model with ${\Omega_{\mathrm{m}}=0.315}$, ${\Omega_{\Lambda}=0.685}$, and ${H_{0}=67.4\,\mathrm{km}\,\mathrm{s}^{-1}\,\mathrm{Mpc}^{-1}}$ \citep{Planck:2018vyg}.


\section{GRB Sample} 
\label{ sec:sample }

As mentioned above, the traditional SGRB/LGRB classification does not always accurately reflect the true physical origin of GRBs. By analogy with SN classifications, \citet{Zhang:2006mb,Zhang:2009uf} proposed that GRBs originating from compact binary mergers and core collapses of massive stars can be named as Type I GRBs and Type II GRBs, respectively, hereafter \mbox{GRB-Is} and GRB-IIs. In this paper, we further classify merger-driven LGRBs (LGRB-Is) as a distinct subgroup within the GRB-I population, particularly in contrast to traditional merger-driven GRBs with relatively short ME durations (SGRB-Is). In Section~\ref{ sec:MP_sample }, we briefly introduce the known SGRB-I and GRB-II samples adopted in this work. Additionally, we describe how we construct the sample of LGRB-Is in Section~\ref{ sec:moLGRB }.


\subsection{SGRB-I and GRB-II samples}
\label{ sec:MP_sample }

In our work, we adopt a sample of 42 SGRB-Is and 273 GRB-IIs with known redshifts from \citet{Minaev:2019unh}, which we  denote as the ``MP Sample'' hereafter. Accordingly, all references to ``SGRB-I'' in this work should be interpreted as \mbox{SGRB-I} within the MP Sample. Collected from 6 experiments in total (Konus-Wind, BeppoSAX, BATSE/CGRO, HETE-2, Swift and Fermi), the MP Sample includes GRB events observed up to January 2019. The MP Sample includes 40 GRB-IIs associated with Type Ic SNe, comprising 19 photometrically confirmed Type Ic SNe and 21 spectroscopically confirmed ones. The most distant SGRB-I event is GRB\,111117A with a redshift of ${z \simeq 2.211}$,\footnote{The spectroscopic redshift measurement for GRB\,111117A is based on its host galaxy's emission lines \citep{Selsing:2017abj}, and the estimated redshift is significantly higher than the previous estimation derived from photometric studies \citep{Margutti:2012aj, 2013ApJ...766...41S}.} while the most distant GRB-II event is GRB\,090423 with a redshift of $z \simeq 8.23$ \citep{Salvaterra:2009ey, Tanvir:2009zz}. Further details regarding this sample and its observed properties can be found in \citet{Minaev:2019unh}.

Building on the intrinsic hardness-total energy correlations \citep{Amati:2002ny}, \citet{Minaev:2019unh} further introduced a new classification indicator for SGRB-Is and GRB-IIs, termed the $\it{EH}$ (`Energy-Hardness') parameter, defined as, 
\begin{equation}
E H = \frac{\left(E_{\mathrm{p}, \mathrm{i}} / 100\,\mathrm{keV}\right)}{\left(E_{\gamma,\mathrm{iso}} / 10^{51}\,\mathrm{erg}\right)^{0.4}} \,,
\label{ eq:EH }
\end{equation}
where ${E_{\mathrm{p}, \mathrm{i}}} = E_{\mathrm{p}} (1+z)$ represents the intrinsic (rest-frame) peak energy of a GRB event, with $E_{\mathrm{p}}$ denoting the position of the extremum (maximum) of the observed time-integrated energy spectrum $\nu F_{\nu}$, and $E_{\gamma,\mathrm{iso}}$ is the isotropic equivalent total energy. Furthermore, \citet{Minaev:2019unh} refined the $\it{EH}$ parameter by incorporating the intrinsic duration, i.e., ${T_{90, \mathrm{i}} = T_{90} / (1+z)}$, where $T_{90}$ is often defined as the time interval during which the integrated photon counts increase from 5\% to 95\% \citep{Kouveliotou:1993yx}.\footnote{It is important to note that the derivation of $E_{\mathrm{p}, \mathrm{i}}$, $E_{\gamma,\mathrm{iso}}$, and $T_{90, \mathrm{i}}$ is significantly influenced by various selection effects, particularly when data from different experiments are considered. Actually, \citet{Minaev:2019unh} have conducted analyses and reliability tests of the $\it{EH}$ and $\it{EHD}$ classification schemes with several selection effetcs. Additional discussions regarding detector-related aspects of our findings are provided in Section~\ref{ sec:discu }.} This leads to a second classification parameter, $\it{EHD}$ (``Energy-Hardness-Duration''),
\begin{equation}
\begin{aligned}
E H D 
& = \frac{E H}{\left(T_{90, \mathrm{i}} / 1\,\mathrm{s}\right)^{0.5}} \\ 
& = \frac{\left(E_{\mathrm{p}, \mathrm{i}} / 100\,\mathrm{keV}\right)}{\left(E_{\gamma,\mathrm{iso}} / 10^{51}\,\mathrm{erg}\right)^{0.4}\left(T_{90, \mathrm{i}} / 1\,\mathrm{s}\right)^{0.5}} \,.
\end{aligned}
\label{ eq:EHD }
\end{equation}
\citet{Minaev:2019unh} suggested that $\mathit{EH} = 3.3$ can be employed for a blind classification of GRB events, and ${\mathit{EHD} =2.6}$ demonstrates superior performance.

\begin{figure*}
    \centering
    \includegraphics[width = 14cm]{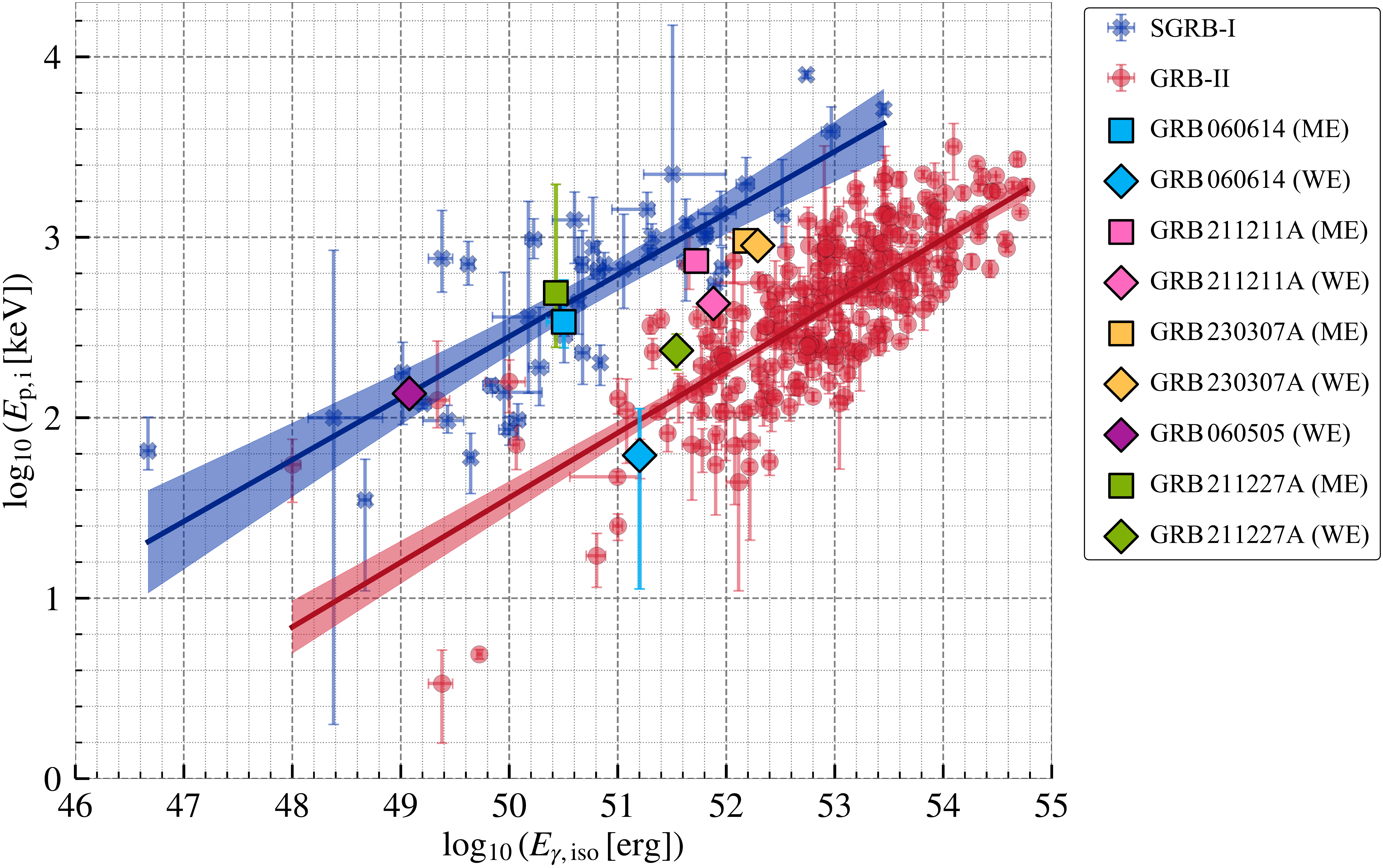}
    \caption{$E_{\mathrm{p},\mathrm{i}} - E_{\gamma,\mathrm{iso}}$ relations for GRBs with known redshifts. All error bars correspond to $1\sigma$ confidence intervals. The solid lines, accompanied by shaded areas, represent the best-fit correlations and the 90\% credible intervals for SGRB-Is (blue) and GRB-IIs (red), respectively. Solid squares and diamonds represent the properties of ME and WE phases for our LGRB-I sample, respectively.} 
    \label{ fig:EP_Eiso }
\end{figure*}


\subsection{LGRB-I Sample}
\label{ sec:moLGRB }

It is believed that GRB\,060614, as well as the more recently detected GRB\,211211A and GRB\,230307A, are three redshift-known LGRB-I events associated with detected KN signals \citep{Yang:2015pha, Jin:2015txa, Troja:2022yya, JWST:2023jqa, 2024Natur.626..717K, Wang:2024ems, 2024Natur.626..742Y}. We collected their observed data and derived properties from the previous studies, listing them in Appendix~\ref{ sec:appenA } (see Table~\ref{ tab:mLGRB }). Additionally, we include two more nearby LGRBs with known redshift $z \lesssim 0.3$ but lacking expected SN observations: GRB\,060505 \citep{Fynbo:2006mw, Ofek:2007kb, Thone:2007ee} and GRB\,211227A \citep{Lu:2022tpu, Ferro:2023rbh},\footnote{Sensitive GRB-associated SN searches are regularly undertaken for GRBs within $z \lesssim 0.3$ \citep{Troja:2022yya}.} which are expected to have merger origins. Notably, except for GRB\,060505, the other LGRB-I events exhibit an EE component following the hard, long-duration ME phase.\footnote{For GRB\,230307A, which is exceptionally bright, the Fermi Gamma-ray Burst Monitor suffers from pulse pile-up and data loss during its ME phase. Spectral analysis typically excludes the post-trigger bad-time interval of $3-7\,\mathrm{s}$ \citep{2023GCN.33551....1D, JWST:2023jqa} and selects specific sodium iodide (na) and bismuth germanate (b1) detectors \citep{Wang:2023yky, Peng:2024nof, Wang:2024ems}.} Given that the EE component may remain undetected in some GRB events, the ME phase can be treated as the whole emission (WE) phase in such cases. Thus, following \citet{Zhu:2022kbt}, we consider ME and WE properties as two independent sets of information for our LGRB-I sample. This treatment serves as a heuristic attempt to explore potential shared properties among LGRB-I events in subsequent analyses.


\section{Empirical Relations for LGRB-Is} 
\label{ sec:corre }

We analyze several GRB classification methodologies in this section, with a more detailed discussion of our newly proposed relation for LGRB-Is in Section~\ref{ sec:discu }.


\begin{figure*}
    \centering
    \includegraphics[width = 14cm]{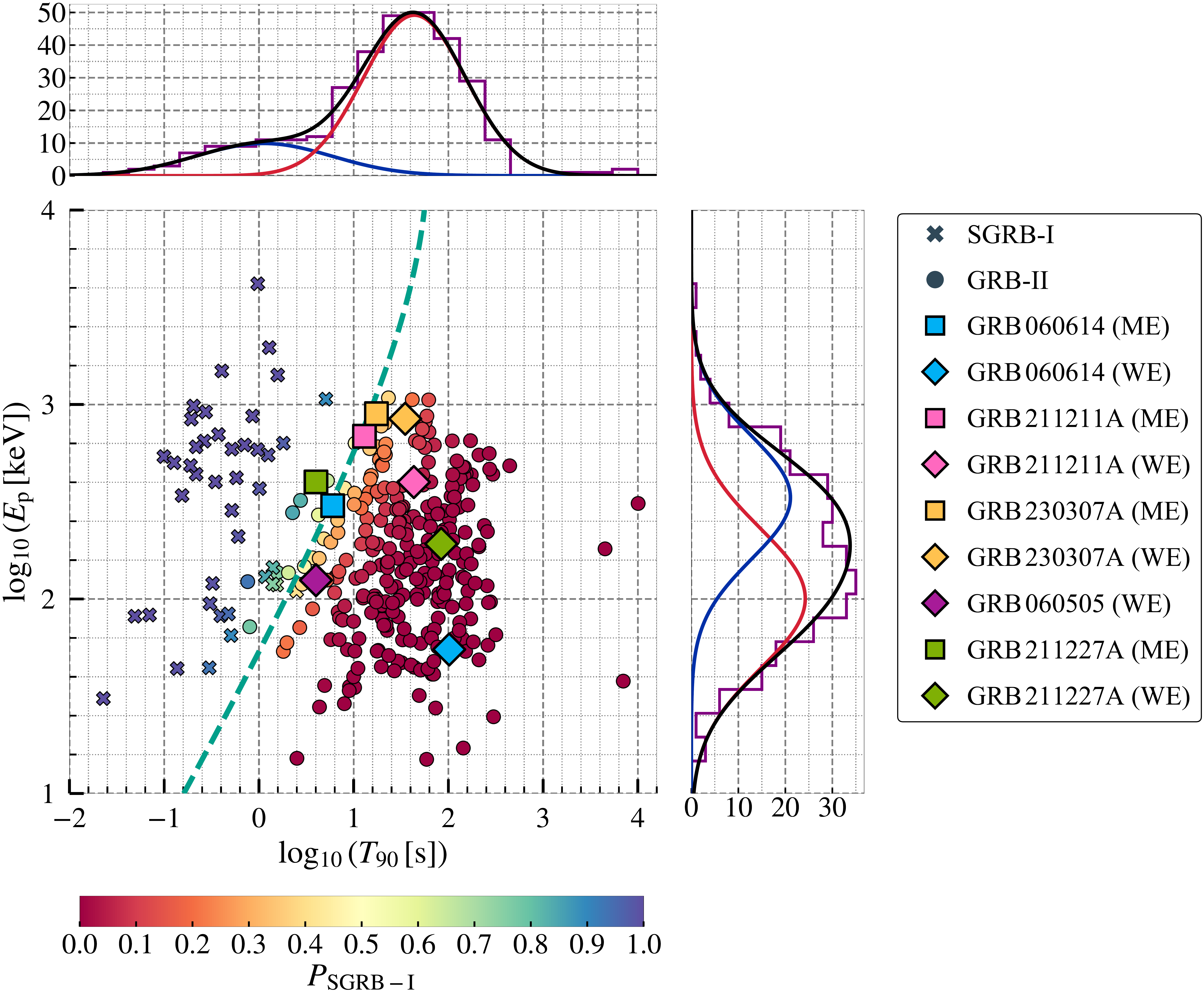}
    \caption{GRB classification diagram in the $E_{\mathrm{p}}-T_{90}$ domain. The crosses and circles represent the known classification of SGRB-Is and GRB-IIs in the MP sample. Based on two independent 2-dimensional Gaussian components, the color scale depicts the probability of a GRB event being a SGRB-I ($P_{\mathrm{SGRB-I}}$). The dashed cyan line indicates the dividing line with $P_{\mathrm{SGRB-I}} = 0.5$. In the top and right projected histograms, the blue and red lines represent the contributions from SGRB-I candidates (smaller $T_{90}$, larger $E_{\mathrm{p}}$) and GRB-II candidates  (larger $T_{90}$, smaller $E_{\mathrm{p}}$), while the black solid lines are the combined distributions of two Gaussian components.} 
    \label{ fig:EP_T90 }
\end{figure*}

\subsection{$E_{\mathrm{p},\mathrm{i}} - E_{\gamma,\mathrm{iso}}$ Relation}
\label{ sec:amati_relation }

We begin by presenting the $E_{\mathrm{p},\mathrm{i}} - E_{\gamma,\mathrm{iso}}$ relations for SGRB-I and GRB-II populations, along with our LGRB-I sample, in Figure~\ref{ fig:EP_Eiso }. It clearly shows that LGRB-Is exhibit a broad range of $E_{\gamma,\mathrm{iso}}$ values spanning several magnitudes, regardless of whether the ME or WE components are considered. Notably, the $E_{\mathrm{p},\mathrm{i}}$ values of the ME components for each LGRB-I event are generally higher than those of the corresponding WE components. From the overall distribution of LGRB-I events, focusing solely on the ME phase suggests a resemblance to SGRB-Is; however, the classification becomes significantly more challenging when WE phases are considered. In other words, Figure~\ref{ fig:EP_Eiso } suggests that no single linear correlation can simultaneously account for both the ME and WE components across all LGRB-I events in the $E_{\mathrm{p},\mathrm{i}} - E_{\gamma,\mathrm{iso}}$ diagram.


\subsection{$E_{\mathrm{p}}-T_{90}$ Relation}
\label{ sec:Ep_T90_relation }

Next, we examine the $E_{\mathrm{p}}-T_{90}$ domain, as shown in Figure~\ref{ fig:EP_T90 }, which is also widely used for GRB classification in the literature \citep{Shahmoradi:2014ira, Gruber:2014iza, vonKienlin:2020xvz, Zhu:2022kbt, Zhu:2024xtm}. By excluding the LGRB-I sample, the $E_{\mathrm{p}}-T_{90}$ distribution is characterized using two independent 2-dimensional Gaussian components, with the histograms of each parameter and their corresponding fits projected onto the top and right panels. The blue and red lines represent the contributions from SGRB-I candidates (smaller $T_{90}$, larger $E_{\mathrm{p}}$) and GRB-II candidates  (larger $T_{90}$, smaller $E_{\mathrm{p}}$), while the black solid lines are the combined distributions of two Gaussian components. The probability of a GRB event being classified as an SGRB-I ($P_{\mathrm{SGRB-I}}$) is calculated by evaluating the GRB's $T_{90}$ and $E_{\mathrm{p}}$ values under such a two-component Gaussian mixture model. Specifically, we compute the joint likelihoods for the SGRB-I and GRB-II components, and define $P_{\mathrm{SGRB-I}}$ as the normalized probability associated with the SGRB-I component. The resulting $P_{\mathrm{SGRB-I}}$ is then represented visually through the color scale in Figure~\ref{ fig:EP_T90 }.

As illustrated in Figure~\ref{ fig:EP_T90 }, most SGRB-Is and GRB-IIs are accurately classified, with $P_{\mathrm{SGRB-I}}$ approaching 1 and 0, respectively. Based on these Gaussian distributions,  we further plot the boundary where $P_{\mathrm{SGRB-I}} = 0.5$ and find that all ME components of LGRB-Is lie near this boundary, indicating the challenge of definitively classifying them as either SGRB-I or GRB-II events. When the WE phases are considered, the distribution of these LGRB-Is deviates significantly from this boundary, showing no universal correlation across both the ME and WE components, similar to that observed in the $E_{\mathrm{p},\mathrm{i}} - E_{\gamma,\mathrm{iso}}$ domain in Figure \ref{ fig:EP_Eiso }.


\subsection{$\it{EH}-T_{\mathrm{90,i}}$ Relation}
\label{ sec:EHD_T90_i_relation }

\begin{figure*}
    \centering
    \includegraphics[width = 14cm]{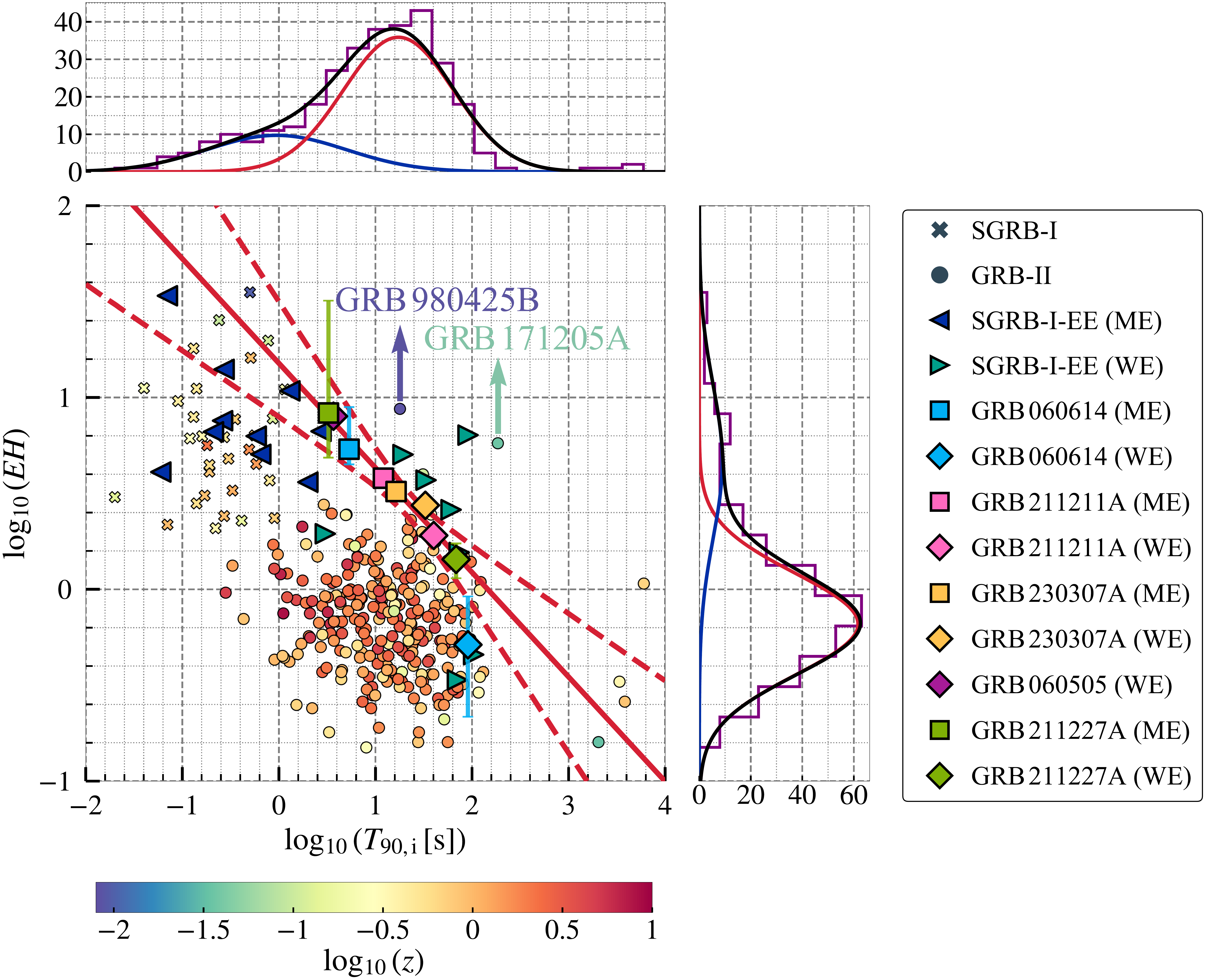}
    \caption{GRB classification diagram in the ${\it{EH}-T_{\mathrm{90,i}}}$ domain. The colorbar now indicates the redshift. The ME and WE components of SGRB-I-EE events are denoted by the left and right triangles, respectively. The best-fit ${\it{EH}-T_{\mathrm{90,i}}}$ correlation is shown as a solid line, with the 90\% credible intervals indicated by dashed lines. The two circular points above the whole GRB-II population correspond to GRB\,980425B ($\mathit{EH} = 8.72$) and GRB\,171205A ($\mathit{EH} = 5.78$).} 
    \label{ fig:EH_T90_with_z }
\end{figure*}

In the ${\it{EH}-T_{\mathrm{90,i}}}$ domain (Figure~\ref{ fig:EH_T90_with_z }), we find that all LGRB-Is, regardless of whether their ME or WE components are considered, probably show a universal correlation in log-log space that remains unexplored. Their distribution appears to occupy a distinct parameter space, serving as a ``bridge'' between the traditional SGRB-I and GRB-II populations. Notably, we do not identify similar correlations within the traditional SGRB-Is and GRB-IIs.

Similar to most LGRB-Is, which are typically identified by an EE phase following the ME phase, we note that an EE component is also observed in some SGRB-I events. Given the similar partially observed properties and the same merger origins between LGRB-Is and SGRB-Is with EE components (\mbox{SGRB-I-EEs}), we then explore and discuss their ${\it{EH}-T_{\mathrm{90,i}}}$ correlations. In the MP Sample, the $E_{\mathrm{p},\mathrm{i}}$ ($E_{\mathrm{p}}$), $E_{\gamma,\mathrm{iso}}$, and $T_{90,\mathrm{i}}$ ($T_{90}$) values for these SGRB-I-EE events are derived exclusively from their ME phases. We thus collect the WE properties of \mbox{SGRB-I-EE} events from the literature and list them in Table~\ref{ tab:mSGRB_wEE }. We first calculate the Pearson correlation coefficient ($r_{\rm P}$) for LGRB-I events in ${\mathrm{log_{10}}(\it{EH})-\mathrm{log_{10}}(T_{\mathrm{90,i}})}$ space, yielding $r_{\rm P} = -0.95$. When additionally including both the ME and EE components of \mbox{SGRB-I-EE} events, we find that the coefficient changes to $r_{\rm P} = -0.72$. This suggests that the distribution between $\mathrm{log_{10}}(\it{EH})$ and $\mathrm{log_{10}}(T_{\mathrm{90,i}})$ for the SGRB-I-EE population exhibits more dispersion and weaker correlation compared to that of the LGRB-I population, which in turn motivates us to explore the probable universal linear correlation between ${\mathrm{log_{10}}(\it{EH})}$ and $\mathrm{log_{10}}(T_{\mathrm{90,i}})$ in our LGRB-I sample.

We then consider that the universal ${\it{EH}-T_{\mathrm{90,i}}}$ correlation among LGRB-Is can be described by a linear model in log-log space, expressed as:
\begin{equation}
\mathrm{log_{10}}(\mathit{EH}) = K\,\mathrm{log_{10}} \left(\frac{T_{90, \mathrm{i}}}{1\,\mathrm{s}}\right) + B\,,
\label{ eq:linear_model }
\end{equation}
where $K$ and $B$ represent the power law index and the constant factor, respectively. By fitting the properties of our LGRB-I events\footnote{Since the $E_{\mathrm{p},\mathrm{i}}$ ($E_{\mathrm{p}}$) value for GRB\,060505 lacks error information, we exclude it from the linear model fitting. However, as clearly shown in Figure~\ref{ fig:EH_T90_with_z }, we note that its position is close to our best-fit line.} listed in Table~\ref{ tab:mLGRB }, we obtain the final results as ${K = -0.55_{-0.10}^{+0.14}}$, and ${B = 1.18_{-0.19}^{+0.15}}$. By incorporating the derived $\it{EH}-T_{\rm 90,i}$ relation with the $\it{EH}$ definition in Equation~(\ref{ eq:EH }), we can further derive a power-law empirical correlation among $E_{\mathrm{p,i}}$, $E_{\gamma,\mathrm{iso}}$, and $T_{\mathrm{90,i}}$ for LGRB-I events,
\begin{equation}
\frac{E_{\mathrm{p, i}}}{100\,\mathrm{keV}} \propto \left(\frac{E_{\gamma,\mathrm{iso}}}{10^{51}\,\mathrm{erg}}\right)^{0.4}\left(\frac{T_{90, \mathrm{i}}}{1\,\mathrm{s}}\right)^{-0.55_{-0.10}^{+0.14}} \,.
\label{ eq:Kang_relation }
\end{equation}
Despite the limited sample size of the collected LGRB-Is, these findings provide valuable insights and motivate further investigation. Although the median value of $EH$ for GRB\,060614 shows a slight deviation from our proposed universal correlation, we note that it also carries relatively large uncertainties, with its upper limit still falling within the 90\% range of the proposed correlation. As more LGRB-I events are observed in the future, these correlations are expected to undergo more rigorous testing, enabling a more comprehensive assessment of their physical significance.


\section{Discussion} 
\label{ sec:discu }

In addition to the universal linear correlation shown in Figure~\ref{ fig:EH_T90_with_z }, the overall distribution of the LGRB-I population in the ${\it{EH}-T_{\rm 90,i}}$ domain is located above that of the GRB-II population,\footnote{We also examined the classification results based on $\it{EHD}$ values (see Table~\ref{ tab:mLGRB }), finding that LGRB-Is commonly show higher $\it{EHD}$ values compared to most GRB-IIs. This is consistent with the results of \citet{Zhu:2024xtm}, which used alternative SGRB-I and GRB-II samples primarily derived from the Fermi catalog.} due to the generally larger $\it{EH}$ values of LGRB-Is compared to most GRB-II events. Given that all LGRB-Is analyzed in this study are relatively nearby events with redshifts of $z \lesssim 0.3$, we check whether our newly proposed correlation could result from the low-$z$ selection effect. This concern arises since \mbox{GRB-IIs} with lower $E_{\gamma,\mathrm{iso}}$ and some low-luminosity \mbox{GRB-IIs} are more easily detected at low redshifts, potentially leading to higher $\it{EH}$ values compared to high-$z$ GRBs with similar durations and hardness. For example, as illustrated in Figure~\ref{ fig:EH_T90_with_z }, some low-$z$, low-luminosity GRB-IIs associated with spectroscopically confirmed SNe, such as GRB\,980425B and GRB\,171205A, lie above the bulk of the GRB-II population. These two events, characterized by $E_{\gamma,\mathrm{iso}} \lesssim 10^{49}\,\mathrm{erg}$, have higher $\it{EH}$ values than most GRB-IIs, further complicating their classification based solely on $\it{EH}$ or $\it{EHD}$ values.

However, as discussed in Section~\ref{ sec:amati_relation }, the $E_{\gamma,\mathrm{iso}}$ values of LGRB-Is exhibit a broad range distribution, regardless of whether the ME or WE components are considered, indicating that our new relation for LGRB-Is cannot be solely attributed to the \mbox{low-$z$} selection effect. Notably, GRB\,211211A and GRB\,230307A are not among the faintest bursts, even when compared to the GRB-IIs population. Moreover, Figure~\ref{ fig:EH_T90_with_z } does not suggest that GRB-IIs further below the best-fit line necessarily correspond to higher redshifts. Some GRB-IIs that deviate significantly from our correlations are located at low redshifts, further undermining a simple relationship between deviation and redshift. We therefore propose that our relations are likely intrinsic, driven by relatively larger $E_{\mathrm{p,i}}$ values of LGRB-Is, rather than the low-$z$ selection effect.\footnote{Reliable redshift measurements are crucial for accurate classification. LGRB-I events would shift toward the central region of the parameter space occupied by GRB-IIs if placed at higher redshifts. Such a scenario could arise if the redshift of a LGRB-I is inferred from a misidentified host galaxy, possibly due to the typically larger offsets from their true host galaxies, or the faintness of the actual hosts renders it challenging to identify \citep{Troja:2022yya}. Consequently, the possibility that some phenomenological LGRBs may have merger origins cannot be ruled out.}

On the other hand, as shown in Figure~\ref{ fig:EH_T90_with_z }, the distributions of both the ME and WE components for different \mbox{SGRB-I-EE} events show larger intrinsic scatter. Since LGRB-I and \mbox{SGRB-I-EE} events share similar phenomenological features, further exploration of the LGRB-I population can provide new insights into the study of merger-origin GRB events.

The introduction of a new empirical relation often inspires efforts to identify the underlying physical mechanisms, though the formulation of an adequate model to explain the remarkably linear relations in Figure~\ref{ fig:EH_T90_with_z } remains a challenge. GRB production and propagation involve numerous complex physical processes that significantly influence the $E_{\mathrm{p}, \mathrm{i}}$, $E_{\gamma,\mathrm{iso}}$, and $T_{90, \mathrm{i}}$ properties across different events \citep{Zhang:2018ond}. Furthermore, the determination of these parameters is heavily affected by various selection biases, especially when the data from different experiments are considered \citep{Qin:2012ht}.\footnote{Nevertheless, taking GRB\,230307A as an example, we have compared results from Fermi and GECAM reports \citep{Sun:2023rbr} and found discrepancies of only $\lesssim 5\%$ in their $\mathrm{log}_{10}(T_{90, \mathrm{i}})$ and $\mathrm{log}_{10}(E_{\mathrm{p}, \mathrm{i}})$ values.} Accurately estimating unbiased GRB properties is virtually impossible, which adds to the complexity of interpreting the correlations we have identified for the ME/WE components of LGRB-Is. This is particularly puzzling when contrasted with the pure clustering behavior observed in SGRB-Is and GRB-IIs populations. We leave these questions for future investigation.


\section{Conclusion} 
\label{ sec:conclu }

In this work, we collect a sample of redshift-known LGRB-I events to investigate their observed properties and explore potential universal correlations. Both the ME and WE phases of these bursts exhibit long durations, yet it is widely accepted that their origins are linked to mergers. Our analysis reveals that LGRB-Is display a broad range of $E_{\gamma,\mathrm{iso}}$, regardless of whether the ME or WE components are considered. When focusing solely on the ME phases of LGRB-Is in the $E_{\mathrm{p},\mathrm{i}} - E_{\gamma,\mathrm{iso}}$ domain, they show a stronger resemblance to SGRB-Is; however, this classification becomes ambiguous when the WE phases are included. We also examine the $E_{\mathrm{p}}-T_{90}$ domain and find that the ME components of all LGRB-I events align closely with the boundary between traditional SGRB-I and GRB-II populations. However, once the WE phases are considered, this alignment becomes less consistent, indicating a distinct shift in the observed trends.

Nevertheless, when we turn our attention to the ${\it{EH}-T_{\mathrm{90,i}}}$ domain, we find that all LGRB-Is, regardless of whether the ME or WE components are considered, show a probable universal linear relationship in log-log space (see Figure~\ref{ fig:EH_T90_with_z }), one that has not been explored in detail before. Their distribution also seems to occupy a distinct parameter space compared to traditional SGRB-Is and GRB-IIs. By incorporating the relations with the definition of $\it{EH}$, we further derive a power-law relationship for LGRB-Is among the three parameters: $E_{\mathrm{p,i}}$, $E_{\gamma,\mathrm{iso}}$, and $T_{\mathrm{90,i}}$ [see Equation~(\ref{ eq:Kang_relation })]. 

The origin and physical mechanism for the proposed universal correlation of LGRB-Is remain an open challenge. This is particularly evident when compared with the relatively weak correlations or lack of correlation observed in SGRB-I with/without EE emissions and GRB-II populations. The observed universality of our relations highlights the necessity for further investigation into the LGRB-I population. Considering the rapid development of both ground-based and space-based gravitational-wave astronomy, future multimessenger observations of the potential associations between GW events and LGRB-Is could provide further insights into the properties of the LGRB-I population \citep{Yin:2023gwc, Kang:2023rux, Moran-Fraile:2023oui, Zhong:2024sry}. 


\begin{acknowledgements}
\label{ sec:acknow }

We would like to thank the anonymous referee for helpful comments and suggestions. We thank R. L. Becerra, Guangxuan Lan, Xishui Tian, Qinyuan Zhang, Jiahang Zhong, and Jinghao Zhang for useful discussions. YK, ZW, and LS are supported by the National SKA Program of China (2020SKA0120300), the Beijing Natural Science Foundation (1242018), and the Max Planck Partner Group Program funded by the Max Planck Society. YK is supported by the China Scholarship Council (CSC). ET and YYH are supported by the European Research Council through the Consolidator grant BHianca (grant agreement ID~101002761). JPZ acknowledges support from the Australian Research Council Centre of Excellence for Gravitational Wave Discovery (OzGrav) through project No. CE17010004.

\end{acknowledgements}


\bibliographystyle{aa}
\bibliography{refs} 


\begin{appendix}

\onecolumn
\section{Tables of the LGRB-I sample and the WE properties of SGRB-I-EE events}
\label{ sec:appenA }

We present the sample of LGRB-I events and their properties in Table~\ref{ tab:mLGRB } to serve the analysis in the main text. The three redshift-known LGRBs, GRB\,060614, GRB\,211211A and GRB\,230307A exhibit KN signals with merger origins. Additionally, we include two nearby ($z \lesssim 0.3$) LGRBs with known redshifts but lacking expected SN observations: GRB\,060505 and GRB\,211227A. Given that the $E_{\mathrm{p},\mathrm{i}}$ ($E_{\mathrm{p}}$), $E_{\gamma,\mathrm{iso}}$, and $T_{90,\mathrm{i}}$ ($T_{90}$) values for SGRB-I-EE events in the MP Sample are derived exclusively from their ME phases, with the EE components excluded, we additionally collect their WE properties and list them in Table~\ref{ tab:mSGRB_wEE }.

\begin{table*}[!h]
    \renewcommand\arraystretch{3.0}
    \centering
    \caption{The sample of LGRB-Is and their properties. The references are: (1) \citet{Yang:2022qmy}; (2) \citet{Peng:2024nof}; (3) \citet{JWST:2023jqa}; (4) \citet{Ofek:2007kb}; (5) \citet{Zhu:2022kbt}.}
    \setlength{\tabcolsep}{0.45cm}
    {\begin{tabular}{c c c c c c c c}
    \toprule 
    \toprule\\
    [-4em]
    GRB    & $z$    & $T_{\mathrm{90,i}}$    & $E_{\gamma,\mathrm{iso}}$    
           & $E_{\mathrm{p,i}}$    & $\it{EH}$    & $\it{EHD}$ & Reference\\
    [-1.5em]
    &        & $[\mathrm{s}]$    & $[10^{51}\,\mathrm{erg}]$    & $[\mathrm{keV}]$\\
    
    \cline{1-8}\\
    [-3.8em]
    060614 (ME)    & 0.125    & 5.33    & $0.32^{+0.02}_{-0.10}$    & $340^{+241}_{-96}$    & 5.37    & 2.33    & (1)\\
    [-1em]
    060614 (WE)    & 0.125    & $90.67^{+4.44}_{-4.44}$    & $1.59^{+0.07}_{-0.13}$    & $62^{+51}_{-51}$    & 0.51    & 0.05    & (1)\\
    [-1em]
    211211A (ME)    & 0.076    & 12.08    & $5.30^{+0.01}_{-0.01}$    & $739^{+13}_{-12}$    & 3.79    & 1.09    & (1)\\
    [-1em]
    211211A (WE)    & 0.076    & $40.13^{+0.06}_{-0.06}$    & $7.61^{+0.11}_{-0.11}$    & $430^{+15}_{-17}$    & 1.91    & 0.30    & (1)\\
    [-1em]
    230307A (ME)    & 0.065    & 16.34    & $15.01^{+0.35}_{-0.35}$    & $956^{+4}_{-3}$    & 3.23    & 0.80    & (2)\\
    [-1em]
    230307A (WE)    & 0.065    & 32.86    & $19.53^{+0.56}_{-0.56}$    & $900^{+4}_{-3}$    & 2.74    & 0.48    & (2, 3)\\
    [-1em]
    060505 (WE)    & 0.0894    & $3.67^{+0.92}_{-0.92}$    & $0.012^{+0.002}_{-0.002}$    & $136^{*}$    & 7.98    & 4.17    & (4)\\
    [-1em]
    211227A (ME)    & 0.228    & 3.26    & $0.269^{+0.025}_{-0.056}$    & $491^{+1474}_{-246}$    & 8.31    & 4.60    & (5)\\
    [-1em]
    211227A (WE)    & 0.228    & 68.40    & $3.48^{+0.16}_{-0.16}$    & $236^{+55}_{-52}$    & 1.43    & 0.17    & (5)   \\
    [0em]
    \bottomrule
    \end{tabular}}
    \begin{tablenotes}
      \item \textbf{Note.} The asterisk marks a $E_{\mathrm{p,i}}$ value obtained from the Swift/BAT Gamma-Ray Burst Catalog (\url{https://swift.gsfc.nasa.gov/results/batgrbcat/index_tables.html}).
    \end{tablenotes}
    \label{ tab:mLGRB }
\end{table*}

\begin{table*}[!ht]
    \renewcommand\arraystretch{2.9}
    \centering
    \caption{The WE properties of SGRB-I-EE events in the MP sample. Note that some events lack well-constrained $E_{\mathrm{p,i}}$ values due to their overly soft EE components. The references are: (1) \citet{Qin:2012ht}; (2) \citet{Villasenor:2005xj}; (3) \citet{Butler:2007hw}; (4) \citet{Krimm:2009nb}; (5) \citet{Zaninoni:2015ooa}; (6) \citet{Zhu:2024xtm}; (7) \citet{Zhang:2020ycl}; (8) \citet{Kagawa:2019hvt}; (9) \citet{AguiFernandez:2021tnw}.}
    \setlength{\tabcolsep}{0.52cm}
    {\begin{tabular}{c c c c c c c c }
    \toprule 
    \toprule\\
    [-4em]
    GRB    & $z$    & $T_{\mathrm{90,i}}$    & $E_{\gamma,\mathrm{iso}}$    
           & $E_{\mathrm{p,i}}$    & $\it{EH}$    & $\it{EHD}$    & Reference\\
    [-1.5em]
    &        & $[\mathrm{s}]$    & $[10^{51}\,\mathrm{erg}]$    & $[\mathrm{keV}]$\\
    \cline{1-8}\\
    [-3.8em]
    
    050709A    & 0.1606    & 123.21    & ---    & ---    & ---    & ---    & (1, 2)\\
    [-1em]
    050724A    & 0.2576    & $73.16^{+1.59}_{-1.59}$    & $0.24^{+0.33}_{-0.03}$    & $88^{+1016}_{-68}$    & 1.56    & 0.18    & (3)\\
    [-1em]
    061006A    & 0.4377    & 90.35    & $2.3^{+0.6}_{-0.6}$    & $888^{+374}_{-207}$    & 6.36    & 0.67    & (4)\\
    [-1em]
    061210A    & 0.4095    & 60.52    & $1.5^{+0.8}_{-0.8}$    & $306^{+439}_{-185}$    & 2.60    & 0.33    & (4)\\
    [-1em]
    070714B    & 0.923    & 33.28    & $13.3^{+3.4}_{-3.4}$    & $1044^{+683}_{-342}$    & 3.71    & 0.64    & (4)\\
    [-1em]
    071227A    & 0.384    & $102.95^{*}$    & $1.2^{+0.5}_{-0.5}$    & $49^{*}$    & 0.46    & 0.05    & (4)\\
    [-1em]
    080123A    & 0.495    & 76.92    & ---    & ---    & ---    & ---    & (5)\\
    [-1em]
    110402A    & 0.854    & $19.23^{+0.79}_{-0.79}$    & $21.8^{+1.32}_{-1.32}$    & $1730^{+516}_{-516}$    & 5.04    & 1.15    & (6)\\
    [-1em]
    150424A    & 0.3    & 70    & 4.45    & $61^{+9}_{-9}$    & 0.34    & 0.04    & (7, 8)\\
    [-1em]
    160410A    & 1.717    & $3.02^{+0.59}_{-0.59}$    & 125.89    & $1346^{+633}_{-633}$    & 1.95    & 1.12    & (7, 9)\\
    [0em]
    \bottomrule
    \end{tabular}}
    \begin{tablenotes}
      \item \textbf{Note.} The asterisk marks a value obtained from the Swift/BAT Gamma-Ray Burst Catalog (\url{https://swift.gsfc.nasa.gov/results/batgrbcat/index_tables.html}).
    \end{tablenotes}
    \label{ tab:mSGRB_wEE }
\end{table*}
	
\end{appendix}


\end{document}